\documentclass[a4paper, secnumarabic, twocolumn, amssymb, nobibnotes, aps]{revtex4-2}

\usepackage{graphicx}
\usepackage{pslatex}
\usepackage{xspace}
\usepackage{amsmath}
\usepackage{color}
\usepackage{booktabs}

\begin{document}

\title{Nongeminate Recombination in Planar and Bulk Heterojunction Organic Solar Cells}

\author{A.~Foertig$^1$}\email{afoertig@physik.uni-wuerzburg.de}
\author{A.~Wagenpfahl$^1$}
\author{T.~Gerbich$^1$}
\author{D.~Cheyns$^2$}
\author{V.~Dyakonov$^{1,3}$}
\author{C.~Deibel$^1$}\email{deibel@physik.uni-wuerzburg.de}

\affiliation{$^1$ Experimental Physics VI, Julius-Maximilians-University of W\"urzburg, D-97074 W\"urzburg, Germany}
\affiliation{$^2$ IMEC v.z.w., Kapeldreef 75, 3001 Leuven, Belgium}
\affiliation{$^3$ Bavarian Center for Applied Energy Research e.V. (ZAE Bayern), D-97074 W\"urzburg, Germany}

\date{July 16, 2012}

\begin{abstract}
We investigate nongeminate recombination in organic solar cells based on copper phthalocyanine (CuPc) and C$_{60}$.  Two device architectures, the planar heterojunction (PHJ) and the bulk heterojunction (BHJ), are directly compared in view of differences in charge carrier decay dynamics. We apply a combination of transient photovoltage (TPV) experiments, yielding the small perturbation charge carrier lifetime, and charge extraction measurements,  providing the charge carrier density. In organic solar cells, charge photogeneration and recombination primarily occur at the donor--acceptor heterointerface. Whereas the BHJ can often be approximated by an effective medium due to rather small scale phase separation, the PHJ has a well defined two-dimensional heterointerface. To study recombination dynamics in PHJ devices most relevant is the charge accumulation at this interface. As from extraction techniques only the spatially averaged carrier concentration can be determined, we derive the charge carrier density at the interface $n_{int}$ from the open circuit voltage. Comparing the experimental results with macroscopic device simulation we discuss the differences of recombination and charge carrier densities in CuPc:C$_{60}$ PHJ and BHJ devices with respect to the device performance. The open circuit voltage of BHJ is larger than for PHJ at low light intensities, but at 0.3 sun the situation is reversed: here, the PHJ can finally take advantage of its generally longer charge carrier lifetimes, as the active recombination region is smaller. 

\textbf{This is the pre-peer reviewed version of the following article:\\
A.~Foertig, A.~Wagenpfahl, T.~Gerbich, D.~Cheyns, V.~Dyakonov, and C.~Deibel.
Nongeminate recombination in planar and bulk heterojunction organic solar cells.
Adv.~Ener.~Mater.~2:1483 (2012). DOI:~10.1002/aenm.201200718}
\end{abstract}

%\pacs{72.80.Le, 73.61.Wp, 72.20.Ee, 72.20.Jv}

\maketitle

\section{Introduction}

Organic solar cells have recently reached 9.2\% power conversion efficiency~\cite{service2011} in  bulk heterojunction (BHJ) device architecture~\cite{sariciftci1992}, in which the active layer is composed of a blend of two organic semiconductors, one acting as electron donor, the other as acceptor. An alternative layer composition is the planar heterojunction, with two adjacent thin layers of donor and acceptor, which was indeed the first organic photovoltaic cell with reasonable efficiency as reported by Tang in 1986~\cite{tang1986}. The PHJ approach, however, has not been able to yield the highest reported efficiencies in recent years, although a notable 5.24\% have been achieved recently for a tetraphenyldibenzoperiflanthene (DBP)/C$_{60}$ bilayer~\cite{hirade2011}. Various research groups have investigated the charge carrier dynamics in BHJ systems in the past \cite{clarke2009, tong2010, baumann2011}. Despite an expected bimolecular charge carrier decay with a quadratic dependence of the recombination rate on charge carrier density, an empirically found apparent order of decay higher than two is often reported and discussed in view of capture and release process of charge carriers into and from trap states, respectively~\cite{nelson2003, foertig2009, thakur2011, etzold2011, kirchartz2011, rauh2012}. The complex donor--acceptor morphology and its influence on the charge carrier dynamics complicate the interpretation of these findings, leading to debates among several research groups concerning the origin of the slow nongeminate recombination.

To gain more insight into the impact of phase separation on charge recombination, we investigated a BHJ and a PHJ solar cell in a direct comparison. Both device types were processed by vacuum deposition of copper phthalocyanine (CuPc) and the fullerene C$_{60}$ within a single assembly procedure. Thus, identical processing parameters---except for the device architecture---allow us to rule out an influence  of differing preparation conditions on recombination dynamics and device performance. The current--voltage characteristics of both device architectures were compared and the transient photovoltage (TPV) technique was applied: as expected, the active layer morphology strongly affects charge carrier recombination and thus carrier lifetime. The fact that only charges near the flat planar donor--acceptor interface are relevant for bimolecular recombination in PHJ devices leads to the problem to account for only these charges experimentally. Thus, we calculated the charge carrier density at the interface $n_{int}$ from the experimental open circuit voltage and compared it to the spatially averaged density $n_{ext}$ as determined by charge extraction (CE) experiments. The resulting charge dynamics for the PHJ device were compared to the BHJ solar cells. For a more detailed consideration, we performed macroscopic device simulations, enabling us to assign differences in the open circuit voltage to the respective device architectures and spatially resolved carrier concentrations. Our findings highlight the impact of the active layer composition on the open circuit voltage. Furthermore, we were able to assign the reduced apparent recombination order of total charge carrier dynamics in PHJ devices to the less complex active layer composition and, hence, reduced charge carrier trapping.
\section{Results}

\begin{figure}[h]
	%\centering
	\includegraphics[width=1.0\linewidth]{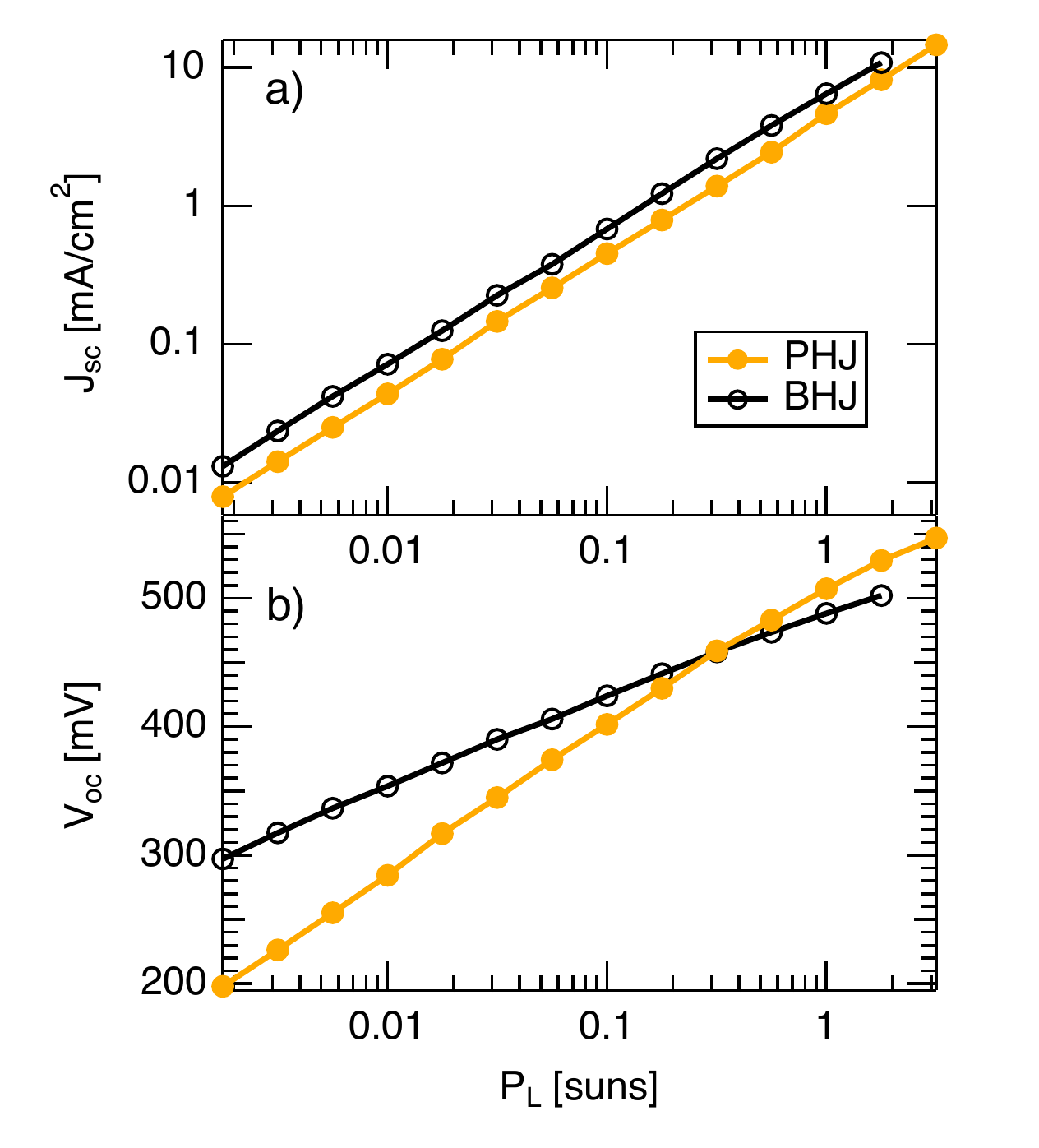}
	\caption{(Color Online) Comparison of a PHJ and a BHJ solar cells based on the materials CuPc and C$_{60}$ at room temperature as a function of bias light: (a) short circuit current density $J_{sc} $ and (b) open circuit voltage $V_{oc}$.}
	\label{fig:iv}
\end{figure}

Several planar and bulk solar cells with the same active layer thickness of overall $d=50$~nm were prepared, all yielding similar I/V characteristics. We processed both CuPc/C$_{60}$ cell types in one run, neglecting to optimize the thickness in view of an increased performance. For the CE and TPV measurements we selected representative samples. The power conversion efficiency of the PHJ was $\eta=1.1\%$, yielding a short circuit current density of $J_{sc}=3.9$~mA/cm$^2$, an open circuit voltage of $V_{oc}=507$~mV and a fill factor of $FF=55$\% at room temperature and 1 sun illumination intensity. Under equivalent conditions the BHJ yielded $\eta=1.4\%$, $J_{sc}=6.5$~mA/cm$^2$, $V_{oc}=488$~mV and $FF=43$\%. In Figure~\ref{fig:iv}, $J_{sc}$ and $V_{oc}$ of PHJ and BHJ solar cells are compared in dependence on incident light intensity. Thereby $J_{sc}$ is significantly higher in BHJ than in PHJ for all light intensities (Fig.~\ref{fig:iv}~(a)). In contrast, we find the open circuit voltage (Fig.~\ref{fig:iv}~(b)) to be lower in the PHJ for low light intensities, but exceeding the voltage of the BHJ for higher light intensities above 0.3~suns. Similar results of $V_{oc}(P_L)$ are shown in Ref.~\cite{xue2005} and discussed with respect to the different ideality factors for both device structures, although without going into detail. As the open circuit voltage is determined by the balance of charge carrier generation and recombination, $G=R$, we focussed on transient photovoltage and charge extraction measurements to study nongeminate recombination under open circuit conditions and further analyze this behavior.

\begin{figure}[h]
	%\centering
	\includegraphics[width=1.0\linewidth]{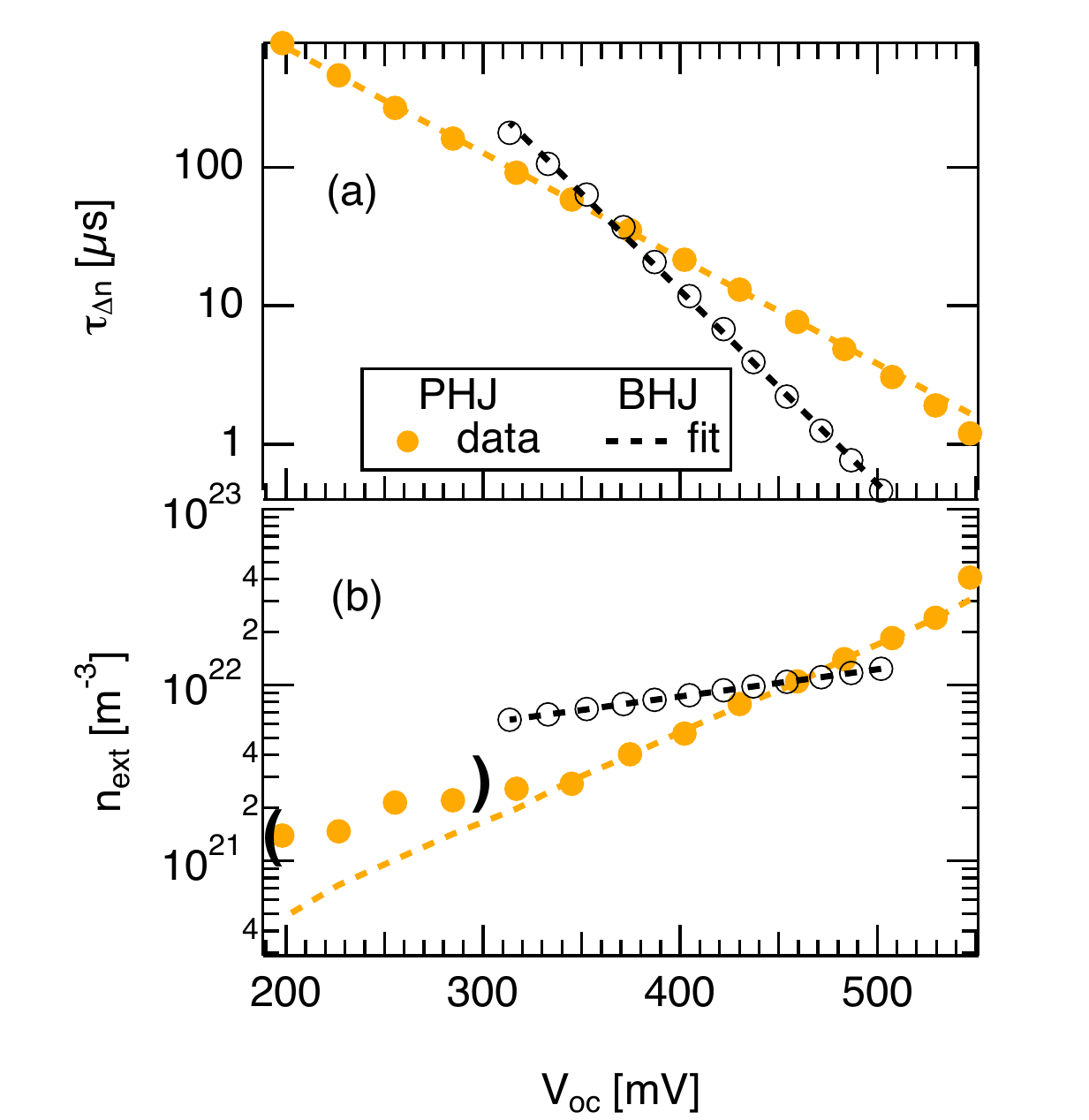}
	\caption{(Color Online) (a) Small perturbation charge carrier lifetime $\tau_{\Delta n}$ in a PHJ (orange) and BHJ (black) device and (b) average charge carrier density determined from CE measurements as a function of the open circuit voltage. Data points within brackets at low light intensities are dominated by capacitive charges and were not accounted for in the fit.}
		\label{fig:tau_voc}
\end{figure}

In Figure~\ref{fig:tau_voc}(a) we present the results of TPV measurements on the PHJ and BHJ device. TPV is based on monitoring the photovoltage decay upon a small optical perturbation during various constant bias light conditions.~\cite{shuttle2008} From the transient  voltage, the small perturbation charge carrier lifetime $\tau_{\Delta n}$ is extracted in dependence of the respective open circuit voltage due to the constant background illumination. Figure~\ref{fig:tau_voc}(a) depicts an exponential dependence of $\tau_{\Delta n}$ on the open circuit voltage $V_{oc}$ similar to TPV studies on polymer:fullerene bulk systems in the past~\cite{shuttle2008, foertig2009, maurano2010}. This finding is valid for both architectures, the present CuPc/C$_{60}$ PHJ and the BHJ device. A steeper slope is observed for the BHJ. The variation of the effective charge carrier lifetime with light intensity indicates a nongeminate loss mechanism for both device architectures.

In order to analyze the charge carrier dynamics, the number of charge carriers participating directly in the recombination process needs to be determined. Within the effective medium approach for BHJ devices, the charge carrier density profile can be approximated to be spatially uniform within the active layer. Spatial variations in the charge carrier generation are compensated to a certain degree by a diffusive current transport. Therefore the charge carrier densities can be measured by a straight forward charge extraction (CE) technique. The situation is different in PHJ devices: polaron generation and recombination are very nonuniform~\cite{cheyns2008}, i.e., localized at the planar donor-acceptor heterointerface. Considering the TPV technique as example, the photovoltage rise upon laser excitation is based on charge carriers successfully generated at the planar donor--acceptor interface from the primary photoexcitations. The subsequent charge carrier decay is proportional to the nongeminate recombination of electron and hole at this interface, not to the overall charge carrier concentration. Thus only charge carriers at the interface ($n_{int}$) are relevant for the charge carrier decay analysis in PHJ solar cells. Nevertheless, for comparison we also performed CE measurements on the PHJ system. As the contribution of capacitive charges is even more dominant for PHJ, the corresponding correction~\cite{shuttle2008b} becomes more important. Therefore an estimation of capacitive charges in the dark and in reverse direction is required and subtracted from the extracted charges determined under illumination in forward direction. The density of extracted charge carriers $n_{ext}$ was calculated assuming in first approximation a constant (uniform) charge carrier distribution in analogy to the BHJ device and appears exponentially dependent on $V_{oc}$ (see Fig.~\ref{fig:tau_voc}~(b)). We point out that our results are in contradiction to the findings of Credgington et al.~\cite{credgington2011}, who found a linear dependence of $n_{ext}$ on $V_{oc}$ on pentacene/C$_{60}$ devices due to the dominating geometric capacitance. The exponential dependence we describe above suggests that the chemical capacitance of the present active material CuPc/C$_{60}$ contributes more to the total amount of charges stored at a certain light intensity. In comparison to the BHJ device, the data derived from the planar device is stronger dependent on $V_{oc}$. As mentioned before, the approximation of a uniform charge distribution in PHJ devices might not be precise enough to describe the decay dynamics exactly. The relevant quantity, the charge carrier concentration at the interface $n_{int}$ cannot be determined directly, e.g.\ by any CE technique. Thus another approach is needed. 

\begin{figure}[h]
	%\centering
	\includegraphics[width=1.0\linewidth]{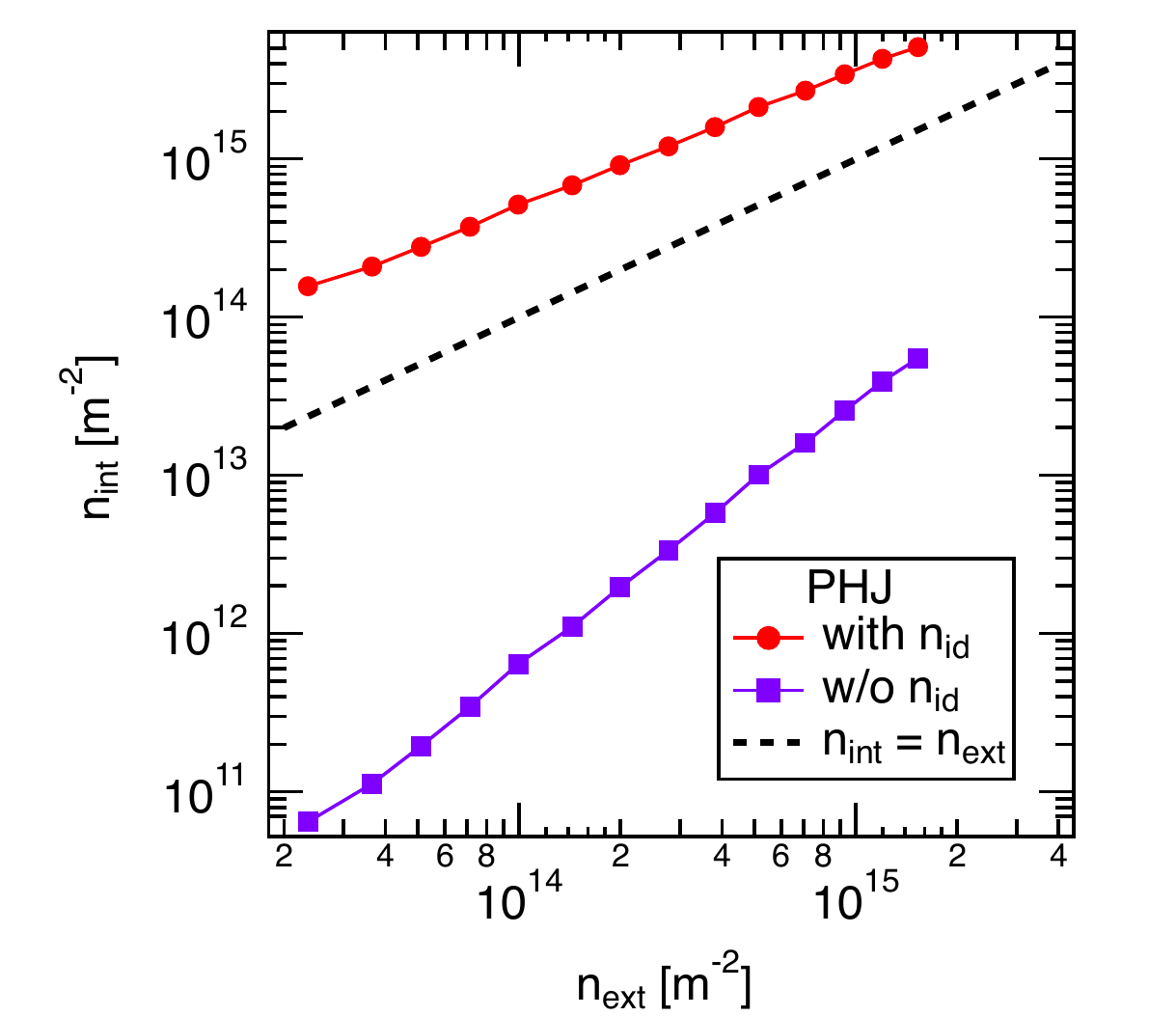}
	\caption{(Color Online) Charge carrier density $n_{int}$ at the donor--acceptor interface calculated using Eqn.~(\ref{eqn:nint}) vs.\ the charge carrier density extracted from CE measurements, $n_{ext}$.  For information, we have also included $n_{int}$ without accounting for the ideality factor $n_{id}$ (by setting it to one). As guide for the eye, the black dashed line with slope 1 indicates a linear dependence, which approximately holds for the calculations including $n_{id}$.}
	\label{fig:ni_n}
\end{figure}

Based on a model introduced in Ref.~\cite{cheyns2008} the open circuit voltage $V_{oc}$ in PHJ solar cells depends---in addition to material constants such as the effective band gap energy $E_g$ and the effective density of states $N_D$---only on the charge carrier concentration $n_{int}$ at the planar interface~\cite{cheyns2008},
\begin{equation}
V_{oc}=\frac{E_g}{q}-\frac{n_{id}kT}{q}\ln\left(\frac{N_D^2}{n_{int}^2}\right).
\label{eqn:voc}
\end{equation}
For simplicity we thereby assumed the effective density of states $N_D$ as well as the interface charge carrier density $n_{int}$ to be equal in donor and acceptor. Furthermore, $q$ is the elementary charge, $kT$ the thermal energy and $n_{id}$ the ideality factor. In Ref.~\cite{cheyns2008} Eq.~\ref{eqn:voc} was derived by considering the position dependent charge concentration and  the related field distribution. Assumptions such as thermionic emission at the contacts and disregard of doping or trapping lead to a simplified expression for $V_{oc}$ in PHJ devices, implicitly representing the charge/field distribution across the device. Within the model, band bending compensates for the energy losses at the contacts.

The effective band gap energy $E_g$ is found to be rather independent of device architecture and was determined from $V_{oc}(T)$ of the BHJ device to be $E_g\approx 1.03$~eV. Thereby $V_{oc}$ turned out to be limited by injection barriers $\Phi$ below T=250~K, roughly estimated as $\Phi\gtrsim 0.2$~eV according to Ref.~\cite{rauh2011}. %The injection barrier $\Phi$ was determined by the difference in y-axis intercept of the linear V_{oc}(T) regime between 300 and 250~K and the data below 250~K.
We used a two-dimensional density of states $N_D=6.3\cdot10^{17}$~m$^{-2}$ at the heterointerface derived from the 3-dimensional density in the macroscopic simulation (see Sec.~\ref{exp}). Accordingly, $n_{int}$ has the unit m$^{-2}$ as well. The ideality factor is determined according to Ref.~\onlinecite{koster2005} from $V_{oc}$ vs $\ln(J_{sc})$ data and determined to be $n_{id}$$\approx $1.94. Eq.~(\ref{eqn:voc}) solved for $n_{int}$ leads to
\begin{equation}
n_{int}=N_D\exp{\frac{(qV_{oc}-E_g)}{2n_{id}kT}}.
\label{eqn:nint}
\end{equation}
We use this equation to calculate the charge carrier density at the planar interface, $n_{int}$,  from the open circuit voltage. In Fig.~(\ref{fig:ni_n}) $n_{int}$ is compared to $n_{ext}$. The latter was determined by CE as described above. We rescaled $n_{ext}$ to a two-dimensional density by multiplying by the active layer thickness. For information, we have also included $n_{int}$ without accounting for the ideality factor $n_{id}$ (by setting it to one).

\begin{figure}[h]
	%\centering
	\includegraphics[width=1.0\linewidth]{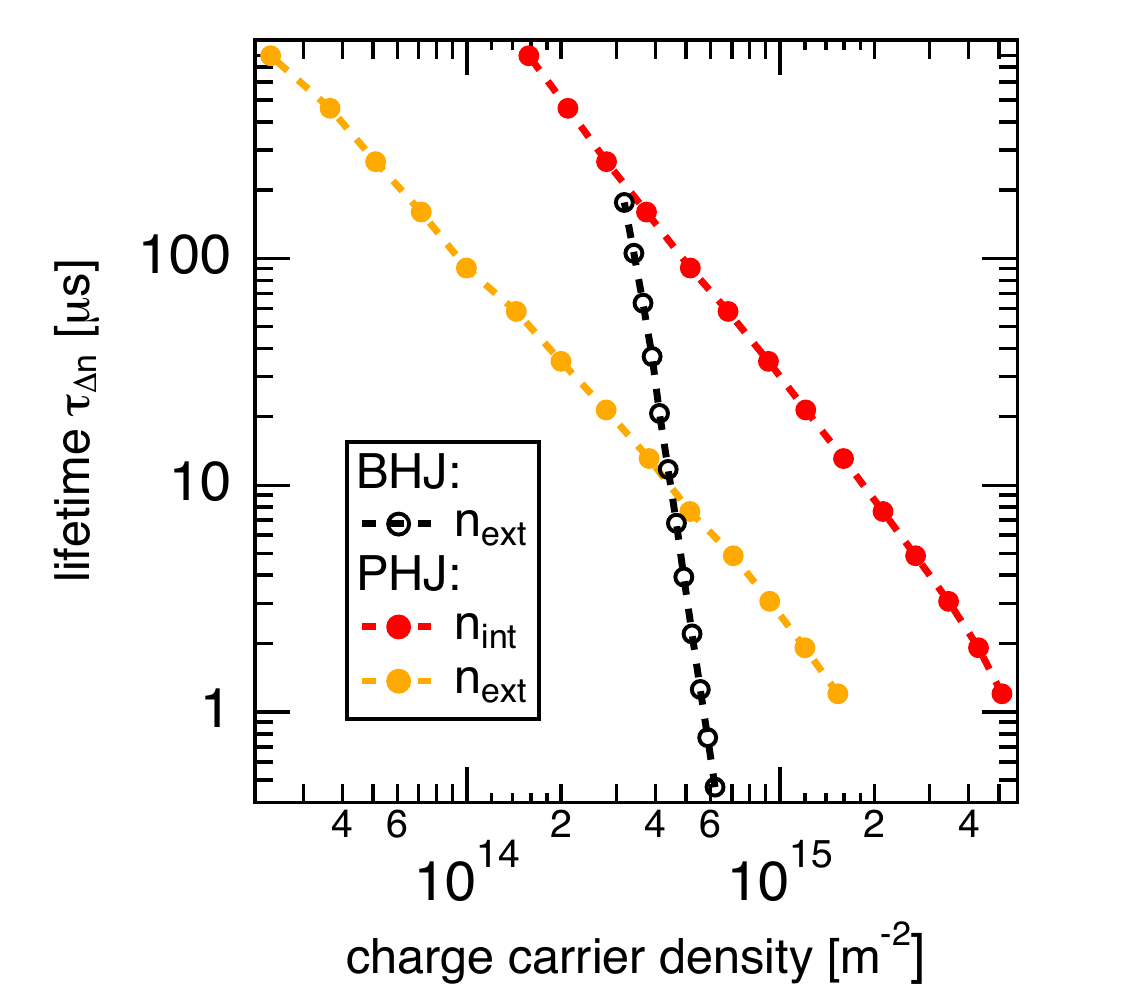}
	\caption{(Color Online) Small perturbation carrier lifetime $\tau_{\Delta n}$ as a function of charge carrier density in two dimensions determined under open circuit conditions. The resulting recombination order is found to depend strongly on the device architecture.}
	\label{fig:tau_ni}
\end{figure}

The respective small perturbation carrier lifetimes $\tau_{\Delta n}$ of PHJ and BHJ in dependence of the appropriate charge carrier density are depicted in Fig.~(\ref{fig:tau_ni}). In order to allow a comparison with the calculated interface charge carrier density $n_{int}$ (Eqn.~(\ref{eqn:nint})), we represented $n_{ext}$ of PHJ and BHJ per area instead of volume, as described above. 

\section{Discussion}

\subsection{Recombination}

The short circuit current density $J_{sc}$ of the CuPc/C$_{60}$ BHJ solar cell is higher than for the PHJ for all light intensities studied, although the dependence on the illumination intensity is similar as shown in Fig.~\ref{fig:iv}.
By assuming a similar absorption in both device architectures, the higher exciton splitting rate in BHJ devices due to the distributed heterointerface results in a higher maximum photocurrent. The charge transport within the spatially disordered photoactive layer of the BHJ device does not seem to limit the current output significantly. Besides a rough estimation of nongeminate charge losses under short circuit conditions according to $J_{sc}\propto P_L^{\alpha}$ reveals negligible losses of about 1.5~\% in the BHJ determined according to Ref.~\onlinecite{koster2011}.

The situation is different for the open circuit voltage, as depicted in Fig.~\ref{fig:iv}~(b). At lower light intensities, $V_{oc_{PHJ}}<V_{oc_{BHJ}}$, whereas from 0.3 suns onward the situation is reversed. Open circuit conditions are equivalent to $G=R$, i.e., the differences in open circuit voltage for PHJ and BHJ can be explained by the different recombination rates.

As pointed out above, in PHJ only the charge carrier density at the heterointerface $n_{int}$ is relevant for the nongeminate recombination process. The almost linear dependence of $n_{int}$ on $n_{ext}$ (Fig.~\ref{fig:ni_n}) suggests the majority of photogenerated charge carriers resides at the planar heterointerface. This implies a strong carrier concentration gradient from the metal--semiconductor interface towards the donor--acceptor interface for low light intensities. However, at higher illumination intensities, the concentration of photogenerated charge carriers increases strongly at the heterointerface, reducing the diffusion current, which ultimately leads to the carrier gradient pointing away from the interface~\cite{cheyns2008}. 

In Fig.~\ref{fig:tau_ni} the small perturbation charge carrier lifetime  $\tau_{\Delta_n}$ in dependence on a two-dimensional charge carrier density of PHJ and BHJ devices are directly compared. Both show a decreasing lifetime with increasing charge carrier density. This behavior is already known from several polymer:fullerene based organic solar cells~\cite{shuttle2008, foertig2009, maurano2010} and can be associated with a dominant bimolecular charge carrier decay. For high charge carrier densities the photogenerated charge carriers in the PHJ have a longer lifetime than in BHJ devices, indicating the strong impact of the device architecture. This behavior can be explained by diffusion of charge carriers away from the heterointerface, which becomes more important in PHJ at high illumination densities due to the photogeneration at this interface. For intense bias light, electrons (holes) diffuse within the acceptor (donor) layer away from the planar heterointerface. Thus, they are screened from bimolecular recombination, which occurs primarily at the donor--acceptor interface, leading to an enhanced charge carrier lifetime in PHJ devices. The latter can be directly related to the open circuit voltage of the PHJ device exceeding the voltage of the BHJ system under high illumination densities (Fig.~\ref{fig:iv}~(b)).
Pointing towards the comparison of $n_{ext}$ and $n_{int}$ in Fig.~\ref{fig:tau_ni} an almost parallel shift with similar slope $\lambda$ becomes evident. The charge carrier density at the interface $n_{int}$, theoretically calculated by Eqn.~(\ref{eqn:nint}) is slightly higher than the experimental parameter $n_{ext}$. The choice of $N_D$, identified from the simulation, thereby comprises the most inaccuracy. Nevertheless a change of $N_D$ would only lead to a parallel shift of $n_{int}$ in Fig.~\ref{fig:tau_ni} without impact on our conclusions. Further a non-linear dependence of the small perturbation charge carrier lifetime $\tau_{\Delta n}$ on the charge carrier density ($\tau_{\Delta n}\propto n^{-\lambda}$) can be derived from Fig.~\ref{fig:tau_ni}. Therby the slope $\lambda$ determines the charge carrier decay order $\lambda$+1 according to $dn/dt\propto n^{\lambda+1}$, usually found to be higher than two if determined by transient experiments on organic solar cells. The deviation from a quadratic decay order is commonly explained by the influence of trapped charge carriers.~\cite{shuttle2008,foertig2009, baumann2011} The data depicted in Fig.~\ref{fig:tau_ni} shows a significantly steeper slope for the BHJ device than for the PHJ device and both devices exceed the quadratic decay order. This finding clearly reveals the influence of the active layer morphology  on charge trapping and suggests a lot more charges being trapped in the complex morphology of a bulk system than in simple PHJ device architecture.

\subsection{Open Circuit Voltage}
\begin{figure}[h]
	%\centering
	\includegraphics[width=1.0\linewidth]{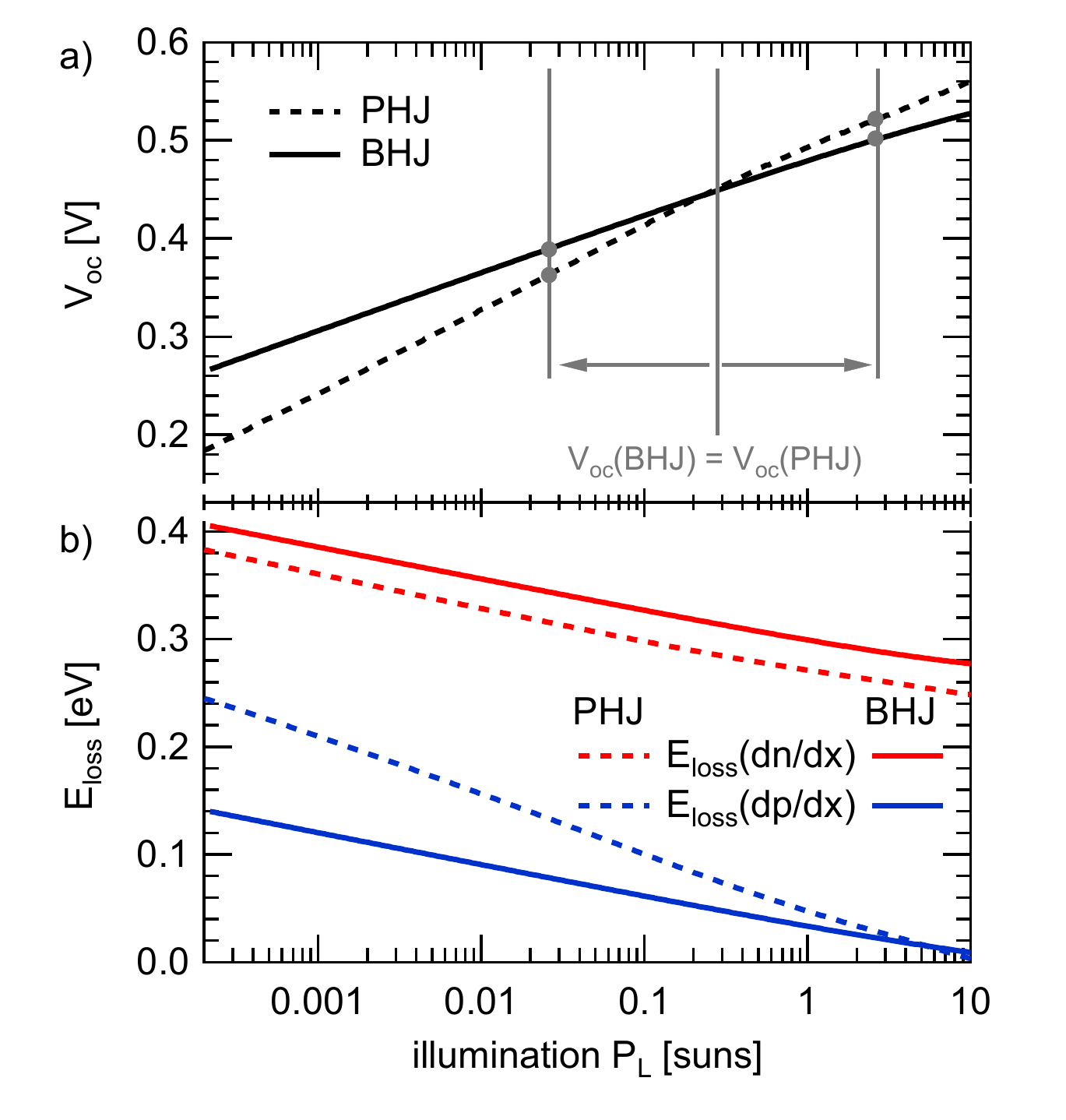}
	\caption{(Color Online) Simulated $V_{oc}$ (a) and potential losses (b) as a function of the illumination in BHJ und PHJ solar cells. The vertical lines at 10~suns before and after the crossing of $V_{oc}$ mark the intensities at which the charge carrier distributions are shown in Fig.~\ref{fig:np}.}
	\label{fig:e_loss}
\end{figure}

We fitted the experimental current--voltage characteristics of BHJ and PHJ devices by our macroscopic device simulation program. This includes an optical transfer-matrix calculation determining the exciton generation profiles as well as an electrical calculation solving the differential equation system of Poisson, continuity and current transport equations \cite{pettersson1999, selberherr1984}. For details we refer to the experimental section. The retrieved parameter set was subsequently used to qualitatively reproduce the experimentally found dependence of $V_{oc}$ on the light intensity $P_L$. The results are depicted in Fig.~\ref{fig:e_loss}~(a) and agree well with the experiment (Fig.~\ref{fig:iv}~(b)). We note that we used experimentally determined parameters as input for the simulation, as described in the experimental section, slightly adjusting them to improve the fits to the measured I/V response (see Supplementary Information). Parameters determined by TPV and CE measurements were not used for the simulation, although experimental and modelled charge carrier concentration and lifetime are qualitatively similar. In the following, our aim is to verify that nongeminate recombination as discussed in the previous section leads to charge carrier distributions which specifically depend on the respective device architecture.  The corresponding loss of electric potential leads to the experimentally observed differences in the dependence of the open circuit voltage on light intensity for PHJ and BHJ devices.

\begin{figure}[h]
	%\centering
	\includegraphics[width=1.0\linewidth]{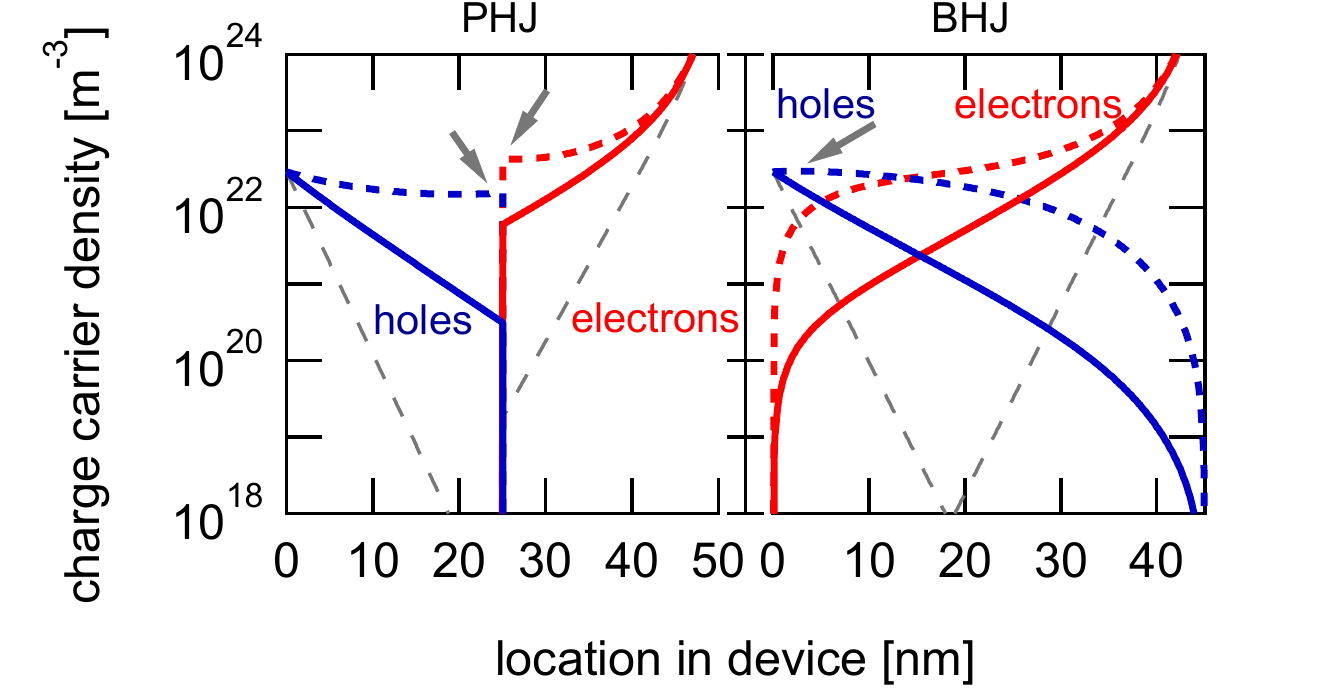}
	\caption{(Color Online) Charge carrier density distribution due to photogeneration and injection for a PHJ (left) and a BHJ (right) device at zero (thin dashed), low (solid) and high (dashed) illumination intensity, corresponding to the leftmost and rightmost markers in Fig.~\ref{fig:e_loss}. With increasing light intensity charges are accumulated at the heterointerface in contrast to the fixed boundary conditions at the electrodes. At high illumination levels space charge regions marked by grey arrows can be observed.}
	\label{fig:np}
\end{figure} 

Mathematically $V_{oc}$ can be described for both device architectures as the effective band gap $E_g$ determined by the lowest unoccupied molecular orbital (LUMO) of C$_{60}$ and the highest occupied molecular orbital (HOMO) of CuPc (cf.~Eqn.~\ref{eqn:voc}), reduced by all potential losses, injection barriers $\Phi$ and Coulomb interaction between charge carriers $E_{loss}$,
\begin{equation}
	V_{oc} = \left(E_{g} - \Phi_n - \Phi_p - E_{loss}\right) / q.
\end{equation} 
In accordance with the experimentally found sum of injection barriers, we set $\Phi_p=0.25$~eV (anode) and $\Phi_n=0.0$~eV (cathode). 

As the net current flow under open circuit condition has to vanish, the sum of the two drift--diffusion equations for electrons and holes can be set to zero and subsequently be written as a function of the electric field. The overall lost electric potential therefore reads
\begin{equation}
	E_{loss}=\int_0^L {\frac{-D_n \frac{\partial n\left(x\right)}{\partial x}+ D_p \frac{\partial p\left(x\right)}{\partial x}}{\mu_n n\left(x\right) + \mu_p p\left(x\right)}}\text{d}x,
	\label{eqn:field}
\end{equation} 
with charge carrier densities $n$, $p$, electron and hole mobilities $\mu_n$, $\mu_p$ as well as their related diffusion constants $D_n$, $D_p$ and the active layer thickness $L$. From Eqn.~\ref{eqn:field}, we find high charge carrier density gradients to be responsible for losses of $V_{oc}$, whereas high charge carrier densities generated by illumination lead to an increase of $V_{oc}$. In order to pinpoint the origin of the detailed $V_{oc}$ vs.\ light intensity dependence, the integral can be separated into two parts for the derivative of electrons and holes, respectively. The results are shown in Fig.~\ref{fig:e_loss}. The corresponding charge carrier profiles for BHJ and PHJ devices are presented in Fig.~\ref{fig:np} for the two light intensities marked in Fig.~\ref{fig:e_loss}~(a), as well as for dark conditions.

As described, the BHJ has a higher open circuit voltage at low light intensities, but due to the lesser slope of $V_{oc}$ vs.\ illumination, as compared to the PHJ, the latter has the higher $V_{oc}$ above 0.3~suns (Fig.~\ref{fig:iv}~(b)). The carrier concentration profiles (Fig.~\ref{fig:np}) show a fundamental difference, since holes and electrons are well separated from each other in PHJ solar cells, but reside within an effective medium in the BHJ. Minority charge carriers in the PHJ, i.e., electrons in the p-conducting as well as holes in the n-conducting material possess a charge carrier density of not more than $10^5$~m$^{-3}$. In contrast, a BHJ solar cell shows a high concentration of both types of charge carriers across the whole extent of the device, which leads to a slightly higher overall generation term than in PHJ.

A metal electrode generally injects electrons as well as holes into a semiconductor described as thermal activation by the thermionic emission theory. Consequently the ratio between injected electrons and holes is an exponential function of the offsets between the metal Fermi level and the LUMO and HOMO levels of the semiconductor. An injection barrier as used here refers to the smaller value of both offsets. For the anode, where holes are the majority charge carriers, this can be stated as
\begin{equation}
	p \propto \exp(-\Phi_p) \quad\text{and}\quad n \propto \exp(-E_{g}+\Phi_p) .
\end{equation}
Hence, in Figure \ref{fig:np} the rightmost and leftmost charge carrier densities, that means at the electrodes, are constant for all illumination levels. Photogenerated charge carriers are accumulated inside the bulk and especially in proximity of the heterointerface.

BHJ devices are described as effective medium with a band gap $E_g$ of 1.05~eV. Therefore the injection of majority carriers at the anode corresponds to hole injection into CuPc, whereas the minority carriers at this interface are electrons injected into C$_{60}$. In contrast, in PHJ devices only one of the two photoactive materials is adjacent to a given electrode, e.g. CuPc (anode). Thus, while the majority (hole) injection barrier corresponds to the BHJ case, the minority carriers have a larger injection barrier of the whole CuPc gap less $\Phi_p$, leading to lower concentrations of minority charge carriers. The situation at the cathode is equivalent, considering that electrons are majorities there. Consequently, the higher a certain injection barrier is, the lower the difference between injected majority and minority charge carriers. 

The charge carrier profiles of BHJ and PHJ device, Fig.~\ref{fig:np}, can be interpreted in terms of  Eqn.~(\ref{eqn:field}). In PHJ devices, the type of majority charge carriers changes always at the heterointerface. In contrast, charge carriers in the BHJ are not restricted to specific layers. Therefore, the spatial position at which the type of majority charge carriers changes from electrons to holes is not constant. Correspondingly, the charge carrier densities and especially their sums are always higher for a BHJ than for a PHJ device. The smaller $V_{oc}$ of the PHJ device in particular at low illumination levels, however, is also caused by the additional steep gradient of the charge carrier concentration. With increasing light intensity, the photogenerated charge at the donor--acceptor interface of PHJ devices has a major impact on the charge carrier gradient within the organic layers. For low light intensities, this gradient leads to charge diffusion towards the interface, while at higher light intensities the gradient can change direction. This is in accordance with longer charge carrier lifetimes derived experimentally by TPV for the planar device (see Fig.~\ref{fig:tau_ni}) as charges diffusing away from the interface are screened from nongeminate recombination. In terms of $V_{oc}$ this weaker gradient corresponds to a stronger reduction of the electric potential loss (Fig.~\ref{fig:e_loss}~(b)) and an open circuit voltage exceeding the one of the BHJ device at intensities above 0.3 suns (Fig.~\ref{fig:e_loss}~(a)). Above one sun, space charges (Fig. \ref{fig:np}, arrows) start piling up in both device architectures. 

In the PHJ device, the space charge is built up at the CuPc/C$_{60}$ interface due to the localised photogeneration, introducing a diffusion of charge carriers away from the heterointerface. For this device geometry, $V_{oc}$ is independent of the injection barriers~\cite{cheyns2008}. In BHJ solar cells, however, the space charge region forms at the anode with its small injection barrier, further lowering the open circuit voltage and potentially also the current into this electrode.

\section{Conclusion}

We studied the recombination mechanisms of planar and bulk heterojunction solar cells. To this end, transient photovoltage and charge extraction techniques were applied to devices based on CuPc and C60. The observed difference in the light intensity dependence of the open circuit voltage between both architectures can be explained by the decay dynamics of charge carriers as well as their spatial distribution. In PHJ, the charge generation primarily occurs at the planar heterojunction, and the photogenerated majority charges are transported in layers with low densities of minority carriers. This induces an enhanced charge carrier lifetime, and carrier profiles that ultimately lead to diffusion away from the heterojunction. In BHJ, however, the complex intermixing of donor and acceptor phase introduces an increased charge carrier recombination. The $V_{oc}$ values for BHJ are higher compared to PHJ at low light intensities. The smaller slope of $V_{oc} (P_L)$ of the former leads to a cross-section, so that at higher light intensities the open circuit voltage of the BHJ remains below the PHJ device. This effect is due to higher recombination rates in BHJ devices as well as the exclusive impact of the injection barrier on the BHJ solar cells. We reproduced the experimental findings, in particular the $V_{oc}$ dependence on light intensity, by a macroscopic device simulation. Thus, this work reveals the impact of the actual charge carrier distribution and charge carrier recombination, both affected by the device architecture, on the open circuit voltage.

\section{Experimental Section}
\label{exp}

All cells studied were prepared at IMEC institute in Belgium and were shipped to Wuerzburg under nitrogen atmosphere for further studies. To ensure no degradation effects I/V characteristics before and after shipment were performed and compared.
Pre-patterned ITO covered glass substrates (kintec, $<$~20~ohm/square) are thoroughly cleaned (soap, deinoized water and solvents). The substrates undergo a 15 minutes UV-ozone treatment prior to the depositions of the organic materials. The organic materials bathocuproine (BCP, Aldrich), copper-phthalocyanine  (CuPc, Aldrich) and fullerene (C$_{60}$ from SES-research) are purified at least ones using thermal gradient vacuum sublimation. The materials are deposited in a high vacuum chamber (base pressure $<1\cdot$10$^{-6}$~Torr) using thermal evaporation, with an evaporation rate around 0.5-1.0~$\text{\r{A}}$/s. Acceptor/donator thickness of the planar device were set to 25/25 nm. The ratio of the blended CuPc/C$_{60}$ layer is kept constant at 1:1. The top contact (Ag) is deposited in the same vacuum chamber, and it defines a cell area of 3~mm$^2$. 

For the TPV measurements the organic solar cells were connected with a 1~G$\Omega$ input resistance of a digital storage oscilloscope. A white high power light emitting diode (Cree) was used to illuminate the active area of the devices, causing a constant rate of generation and recombination. With a Nd:YAG laser pulse (532~nm excitation wavelength, pulse duration 100~ps) an additional generation rate was induced which caused a small perturbation of the photovoltage. The amount of photogenerated charges was derived by charge extraction transients and experimental data was treated as reported in Ref.~\cite{rauh2011}. Densities per area were calculated by considering the active layer area.

The simulations were performed by our organic solar cell simulation software solving the differential equation system of Poisson, continuity and drift--diffusion equations~\cite{wagenpfahl2010}. The light excitation profile was calculated by a transfer matrix method generating excitons in a pure semiconductor respectively polaron pairs in blended materials~\cite{pettersson1999}. The illumination was set to an AM1.5 spectrum. If an exciton diffuses to a heterointerface within its lifetime it gets converted to a polaron pair. The following generation and recombination dynamics were described by the Onsager--Braun and Langevin theory~\cite{onsager1938, langevin1903}.

By fitting the current--voltage characteristics of a PHJ and a BHJ device under illumination and in the dark, we acquired our simulation parameters in within the experimental validated regime. The ITO/BCP and silver electrodes were assigned to work functions of -4.848~eV and -4.05~eV. Respectively the LUMO and HOMO levels of CuPc were set to -3.0~eV and -5.1~eV, those of C$_{60}$ to -4.05~eV and -6.15~eV in accordance with Ref.~\onlinecite{rand2007}.  An effective blend was created combining the CuPc's HOMO with the C$_{60}$'s LUMO level. The charge carrier mobilities for electrons and holes were fitted to 1.7$\cdot$10$^{-8}$~m$^2$/Vs in CuPc, 3$\cdot$10$^{-6}$~m$^2$/Vs in C$_{60}$ and 3$\cdot$10$^{-9}$~m$^2$/Vs in the blended phase. Furthermore the effective density of states was set to 5$\cdot$10$^{26}$~m$^{-3}$ in all semiconductors. The relative dielectric constant was set to 3.4 in CuPc and in the blend, in contrast to 4.0 in pure C$_{60}$. Also the thickness of the BHJ active layer was reduced from 50~nm to 45~nm, still within the preparation error. Within the Onsager--Braun model, a polaron pair distance of 1.2~nm and a decay rate of 1.4$\cdot$10$^5$~s$^{-1}$ were assumed. The absolute value for the decay rate should not be overestimated, as it includes microscopic instead of directly measurable parameters, e.g. high local mobilities required for effective charge separation \cite{veldman2009, deibel2009}. The exciton diffusion length was set to a value larger than the spatial extent of the active layers.

\begin{acknowledgments}
A.F.'s work was financed by the the Dephotex Project (Grant No.~214459) in the 7th framework programme of the European Commission. C.D.\ gratefully acknowledges the support of the Bavarian Academy of Sciences and Humanities. V.D.'s work at the ZAE Bayern is financed by the Bavarian Ministry of Economic Affairs, Infrastructure, Transport and Technology.
\end{acknowledgments}

\end{document}